\documentclass{article}

\usepackage{arxiv}

\usepackage[utf8]{inputenc} 
\usepackage[T1]{fontenc}    
\usepackage{hyperref}       
\usepackage{url}            
\usepackage{booktabs}       
\usepackage{amsfonts}       
\usepackage{nicefrac}       
\usepackage{microtype}      
\usepackage{graphicx}
\usepackage{doi}
\usepackage{amssymb}
\usepackage{latexsym}
\usepackage{epsfig}
\usepackage{amsmath}
\usepackage{multirow}

\def\convinlaw{\stackrel{{\cal L}}{\Longrightarrow }}
\def\convinp{\stackrel{P}{\longrightarrow }}

\def\tends{\rightarrow}

\def\RR{\mathbb R}
\def\ZZ{\mathbb Z}

\def\EE{\mathbb E}

\def\Conj{\mathcal C}

\newtheorem{Theorem}{Theorem}

\newtheorem{Corollary}{Corollary}
\newtheorem{Proposition}{Proposition}
\newtheorem{Definition}{Definition}

\newtheorem{Remark}{Remark}[section]

\title{ Skip-sampling: subsampling in the frequency domain   }

\author{ Tucker S. McElroy\\U.S. Census Bureau \\
  4600 Silver Hill Road, Washington, DC 20233\\
\texttt{tucker.s.mcelroy@census.gov} \And 
       Dimitris N. Politis \\Department of
Mathematics and Halicioglu Data Science Institute  \\
University of California at San Diego \\ La Jolla, CA 92093-0112, USA
      \texttt{dpolitis@ucsd.edu} }

\date{}

\hypersetup{
pdftitle={Skip-sampling: subsampling in the frequency domain },
pdfsubject={},
pdfauthor={Tucker S.~McElroy, Dimitris N.~Politis},
pdfkeywords={Discrete Fourier Transform, Spectral Density, Time Series},
}

\begin{document}
\maketitle

 \begin{abstract}  Over the last 35 years, several bootstrap methods for
time series have been proposed.
Popular `time-domain' methods include
the block-bootstrap, the stationary bootstrap, 
the linear process bootstrap, etc.; subsampling for time series
 is also available, and is closely related to the block-bootstrap.
   `Frequency-domain' bootstrap has been  performed
either by  resampling   the periodogram ordinates
or by resampling   the ordinates of the  Discrete Fourier Transform (DFT).
The paper at hand  proposes a novel construction of 
{\it subsampling} the DFT ordinates, and investigates its 
theoretical properties and realm of applicability.
\end{abstract}



\section{Introduction}
  
Efron (1979) developed the bootstrap for independent and identically
distributed (i.i.d.) data, and paved the way for practical 
nonparametric statistics in the modern era. Soon after,
practitioners were  able 
to apply resampling ideas in a variety of non-i.i.d. situations including the interesting case of dependent data.

Time series analysis has a `time-domain' and a `frequency-domain' 
aspect to it. Consequently, proposals for resampling time series
can be in either of these two flavors. 
There have been several proposals
with regards to `time-domain' resampling plans; these include
the block-bootstrap in its many variations, the stationary bootstrap, 
the linear process bootstrap, etc.; see Chapter 12 of McElroy  and   Politis  (2020) for a description. 
Subsampling for time series
 is closely related to the block-bootstrap; see Politis et al. (1999).
For book-length 
treatments of the state-of-the-art in resampling time series,
see Lahiri (2003a) or Kreiss and Paparoditis (2023).
   
One of the first  papers  on a  `frequency-domain' bootstrap 
was by Franke and H\"ardle (1992),  who proposed   resampling the periodogram ordinates.
  The motivation behind that approach is that  
periodogram ordinates at different Fourier  frequencies are approximately independent.  Different avenues based on this approach 
were pursued by  several researchers including 
  Janas and Dahlhaus (1994), Dahlhaus and Janas (1996), 
 Kreiss and Paparoditis (2003), and Meyer et al. (2020).   

Interestingly, there was an   earlier---albeit unpublished---report by 
Hurvich and Zeger (1987),  who proposed resampling 
the ordinates of the  Discrete Fourier Transform (DFT), as they 
are also approximately independent. Actually, the
aforementioned (approximate)  independence
of periodogram ordinates is a consequence of the 
(approximate) independence  of DFT
ordinates,  since the periodogram is a function of the DFT. 
Hence, resampling the DFT can be thought of as a 
more fundamental  construction; a rigorous
development can be found in Kirch and  Politis (2011). 

Since resampling the DFT is a fundamental  construction, 
the question of possibly {\it subsampling} the DFT presents itself;
this is the subject of the paper at hand.
The basic idea is to divide the DFT (based  on a sample of size $T$)
   into $q$ vectors of length $b$, each consisting of the DFT 
ordinates at frequencies
 separated by $q/T$.  Each such vector  is asymptotically independent of one another, and distributed as  
 a DFT vector based on a sample of size $b$.   
If the statistic at hand is computable based on the DFT alone, it  
could   be re-computed on the smaller DFTs, and an empirical distribution of such
subsample statistics, appropriately centered and normalized, 
 would estimate the original statistic's sampling distribution.

The above construction will  be termed {\it skip-sampling} of the DFT because of the process of skipping over 
some frequencies in putting together the subsample DFT vectors.
The following section gives 
the precise construction as well as 
some background on the properties 
of the DFT. Some theoretical results on skip-sampling  of the DFT
are given in Section \ref{se:Skip-sampling}, while 
 applications to spectral means and ratio statistics 
are given in Section \ref{se:Statistics defined in the frequency domain}.

\section{Problem setup}
 Let $X_1, X_2, \ldots, X_T $ be an observed sample 
from a  strictly stationary time series $\{ X_t \}$, and denote
${\bf X} = {[ X_1, X_2, \ldots, X_T ]}^{\prime}$.
Time series analysis in the `frequency domain' hinges on 
the Discrete Fourier Transform (DFT), which maps the data vector ${\bf X}$
to a vector with (approximately) independent entries. To define the DFT, consider 
the    Fourier frequencies    $\lambda_{\ell}= 2 \pi \ell/T$,
where $\ell$ is an integer satisfying  $[T/2] - T +1 \leq \ell \leq [T/2]$
and $[\cdot]$ denotes integer part;
  this index range corresponds to $-[T/2] \leq \ell \leq [T/2]$ when $T$ is odd, 
   or $-[T/2] +1 \leq \ell \leq [T/2]$ when $T$ is even.\footnote{Some authors
define the DFT using the Fourier frequencies    $\lambda_{\ell}= 2 \pi \ell/T$
for $\ell=0,1,\ldots, T-1$; due to the periodicity present in the DFT, there is
no discrepancy here, other than a re-ordering of the entries of the DFT vector
which will help us exploit some symmetry properties.}
%
   %
      Define a $T \times T$ matrix   $Q$ with complex-valued entries
$  Q_{j k} = T^{-1/2} \, e^{ i j \lambda_{[T/2]-T +k} }$
 for $1 \leq j,k \leq T$.  Note that  $Q$ is unitary, i.e., $Q^{-1} = Q^*$,  the conjugate transpose.  
 The DFT vector  (see Proposition of 7.2.7 of McElroy and Politis (2020) for more details)  is defined as
$  \widetilde{{\bf X}} = Q^* \, {\bf X}$,
which means that the $k$th component of the DFT vector is
 \begin{equation}
 \label{eq:dft-explicit}
   \widetilde{X}_j = T^{-1/2} \sum_{k=1}^T e^{-i k \lambda_{[T/2]-T+j} } X_k.
 \end{equation}
The DFT map is invertible, because
 clearly ${\bf X} = Q \,  \widetilde{{\bf X}}$.  

\subsection{DFT Symmetries}
The DFT vector has certain symmetries; in order to describe these symmetries,
 we define a transposition matrix $P$ (or $P_T$, when we need to annotate its dimension),
 which is a   $T \times T$ matrix with ones on the trans-diagonal
(and zeros elsewhere); its  action on a vector
  is to reverse the order of its components.  
 %
 %
  Let $1_T $ denote the identity matrix of dimension $T$, and
     let $\Pi$ be the permutation matrix that when applied to a column vector  shifts all the components upwards one
   position, and sends the first component to the last (bottom) position.
  The complex conjugate of $z$ is denoted $\Conj z$,
   and the real and imaginary parts   are $\Re z$ and $\Im z$ respectively.
   
\begin{Proposition}
\label{prop:dft-sym}
 If $T$ is odd, the DFT vector  $\widetilde{{\bf X}}$ satisfies
  \begin{equation}
  \label{eq:sym.property-odd}   
    P \, \widetilde{\bf X} = \Conj {\widetilde{\bf X}}.
\end{equation} 
If $T$ is even, the DFT vector satisfies
 \begin{equation}
  \label{eq:sym.property-even}   
  \Pi \,  P \, \widetilde{\bf X} =  \Conj {\widetilde{\bf X}}.
\end{equation} 
\end{Proposition}

\paragraph{Proof of Proposition \ref{prop:dft-sym}.}

First consider the case that $T$ is odd, so $T = 2m+1$ for some
integer $m$.  Then $P \ \widetilde{\bf X}$
 has   entries in reverse order, so that the middle component is
 unchanged but all others are flipped.   Using  (\ref{eq:dft-explicit}),
 $ \widetilde{{X}}_j = T^{-1/2} \sum_{k=1}^T e^{-i k \lambda_{-m-1+j} } X_k$.
 On the other hand,   the $j$th component of $P \ \widetilde{\bf X}$ is  
 $ \widetilde{ {X}}_{T+1-j} = T^{-1/2} \sum_{k=1}^T e^{-i k \lambda_{m+1-j} } X_k$,
  which is the conjugate of  $ \widetilde{X}_j$.  
This proves (\ref{eq:sym.property-odd}).   Next, suppose that $T$ is even,
 so $T = 2m$ for some integer $m$.   For $1 \leq j \leq T-1$, the $j$th component
  of   $ \Pi \,  P \, \widetilde{\bf X}$ is the $j+1$th component of $ P \, \widetilde{\bf X}$,
  which is
\[  
  \widetilde{X}_{T-j} =  T^{-1/2} \sum_{k=1}^T e^{-i k \lambda_{m-j} } X_k
  =  T^{-1/2} \sum_{k=1}^T e^{i k \lambda_{-m+j} } X_k = \Conj \widetilde{X}_j.
\]
 Moreover,  the  $T$th component   of   $ \Pi \,  P \, \widetilde{\bf X}$ is 
 the first component of $P \, \widetilde{\bf X}$, i.e.,  $\widetilde{X}_T$.
  Because $\lambda_m = 2 \pi m/T = \pi$, this number is real, and   hence 
   $\widetilde{X}_T = \Conj  \widetilde{X}_T$.      This proves (\ref{eq:sym.property-even}).  $\quad \Box$
      
     \vspace{.5cm}
     
     In view of Proposition \ref{prop:dft-sym}, we say that the DFT vector 
     satisfies a ``Symmetry Property,'' defined as follows.
     
\begin{Definition}
\label{def:sym-property}
 A length $T$ complex vector $\widetilde{\bf X}$ satisfies the {\it Symmetry Property} if and only if
   (\ref{eq:sym.property-odd}) holds when $T$ is odd and  (\ref{eq:sym.property-even}) holds
   when $T$ is even.
\end{Definition}

 A general length $T$ complex vector ${\bf Z}$ may   satisfy the Symmetry Property, in which case
 necessarily the vector has a particular structure.  If 
         ${\bf Z}$ satisfies (\ref{eq:sym.property-odd}),  it must be the case that
   \[
     \Re [ P {\bf Z}] = \Re [  {\bf Z}] , \qquad \Im [P  {\bf Z}] = - \Im [   {\bf Z}].
     \]
  Because the middle entry of an odd-length $ {\bf Z}$  has its value unchanged
  after application of $P$,  its value must be real.    Similarly, a vector ${\bf Z}$ satisfying  (\ref{eq:sym.property-even})
  has real entries for components $T/2$ and $T$, whereas the subvector of components $1$ through $T/2-1$ is
   the complex conjugate of the transposition of the subvector for components $T/2 +1 $ through $T-1$.
   In this paper we will be constructing DFT vectors in various ways, but we need to ensure that these constructions
  have the correct properties.  In particular, given a complex vector $ {\bf Z}$ it behooves us to know how it
   can be modified so as to have the Symmetry Property.    The following proposition justifies this motivation.
   
\begin{Proposition}
\label{prop:sym-invert}
  If $ {\bf Z}$ has the Symmetry Property, then $Q  {\bf Z}$ has real-valued entries.
\end{Proposition}

\paragraph{Proof of Proposition \ref{prop:sym-invert}.}
 First, noting that $P_{jk} = 1_{ \{ j+k = T+1 \} }$, we obtain
 \begin{align*}
   { \{  \Conj Q  P \} }_{jk} & = \sum_{\ell=1}^T \Conj {Q}_{j \ell}  \, P_{\ell k}
     = \overline{Q}_{j, T+1-k}  \\
     & = T^{-1/2} \, \exp \{ -2 \pi i j ([T/2] - T + T+1 -k ) / T \} \\
          & = T^{-1/2} \, \exp \{ 2 \pi i j (k-1 - [T/2] ) / T \}.
\end{align*}     
  If $T$ is odd, then $-1 - [T/2] = [T/2] - T$,  and 
   $ { \{  \Conj {Q} P \} }_{jk} = Q_{jk}$.  But if $T$ is even, then $-[T/2] = [T/2] - T$ and
    $ { \{  \Conj {Q} P \} }_{jk} = Q_{j,k-1}$ for $2 \leq k \leq T$; also
     $ { \{  \Conj {Q} P \} }_{j 1} = Q_{j,T}$ because
     $\exp \{ 2 \pi ij ([T/2] - T)/T  \} = \exp \{ 2 \pi ij ([T/2] )/T  \}$.
             Therefore,  when $T$ is odd
    $ \Conj {Q} P = Q$, but when $T$ is even $\Conj {Q} P = Q \Pi$.
    Next, because $P$ is idempotent 
\[
  \Conj [ Q \, {\bf Z} ] =  \Conj [Q] P  \, P  \Conj [ {\bf Z} ]
  = \begin{cases}  Q \, P  \Conj [ {\bf Z} ] \quad \mbox{if} \; T \; \mbox{is odd} \\
  	Q \, \Pi    P  \Conj [ {\bf Z} ]  \quad    \mbox{if} \; T \; \mbox{is even}
  	\end{cases}
  	=  Q  {\bf Z},
\]
 using  (\ref{eq:sym.property-odd}) and     (\ref{eq:sym.property-even}).
  Hence $Q {\bf Z}$ equals its own conjugate, and therefore must be real.  $\quad \Box$

\begin{Remark} \rm
\label{rem:symmetrizing} 
  As an application,  we can alter a given complex vector ${\bf Z}$ to have the 
  Symmetry Property as follows.   If the length $T$ is odd, replace the first $[T/2]$
   entries with the conjugate of the first $[T/2]$ entries of $P {\bf Z}$, and
   discard the imaginary part of the middle entry in position $[T/2]+1$.  If the length of $T$ is even,
 we replace components $1$ through $T/2 -1$ with the conjugate of components
  $T-1$ through $T/2 + 1$ (so their order is flipped);  also, the imaginary portions of
  components $T/2$ and $T$ are discarded.
   These operations ensure that the modified ${\bf Z}$ has the Symmetry Property.
\end{Remark}  
 
\subsection{Subsampling   the DFT}
 We now provide details on the novel construction that is at the heart of this paper's methodology.
For simplicity, consider positive integers $q$ and $b$ 
such that\footnote{In practical applications, if such an exact choice is not feasible,
 then one could let $q=[ T/b ]$, and work as if the  data 
were just $X_1, \ldots, X_{bq}$.  }
 $b \, q = T$, and define sub-components of the DFT vector by  
\[
  \widetilde{{\bf X}}^{(j)} = { [ \widetilde{ {X}}_j,  \widetilde{ {X}}_{q+j}, 
	\ldots,  \widetilde{ {X}}_{(b-1)q+j} ]}^{\prime}
\]
 for $  j =1, \ldots,  q$.  In terms of the entire DFT, this operation can be expressed as
$ \widetilde{ {\bf X}}^{(j)} = [  1_b \otimes e_j^{\prime} ] \, \widetilde{{\bf X}}$,
 with $e_j$ the $j$th unit vector in $\RR^q$ and $\otimes$ is the Kronecker product.
Because of the construction of keeping every $q$th  Fourier frequency
and  skipping over the intervening ones, this operation can be called {\it skip-sampling}
 on the DFT, and
$ \widetilde{ {\bf X}}^{(j)}$  is called the $j$th skip-sample DFT; it is
a complex vector of length $b$,  obtained by evaluating the DFT only at
 frequencies of the form  $2 \pi (\ell q + j) /T$, where  $\ell = [b/2]-b+1,\ldots,[b/2]$. 

Recall that the DFT $\widetilde{{\bf X}}$ contains all the information
carried in the sample  $ {\bf X},$ since we can re-create
  $ {\bf X} $ as $ Q \,  \widetilde{ {\bf X}}$. However, 
the $j$th skip-sample DFT $ \widetilde{{\bf X}}^{(j)}$ contains only
a part of the   information carried by the sample  ${\bf X}$;
putting all the skip-sample DFTs $ \widetilde{{\bf X}}^{(j)}$ together
for $  j =1, \ldots,  q$,  we can capture the whole information again. 
In this sense, working with the skip-sample DFTs $ \widetilde{{\bf X}}^{(j)}$ 
for $  j =1, \ldots,  q$ can be considered a form of {\it subsampling in the
frequency domain}; this should be contrasted to  the usual 
  subsampling of a time series in the
{\it time  domain} which is done by carving the sample
$X_1, \ldots, X_T$ into smaller blocks, each 
consisting of $b$ consecutive data points---see Politis et al. (1999). 

Note that  $ \widetilde{{\bf X}}^{(j)}$  will not necessarily 
  have the Symmetry Property; employing the techniques of Remark \ref{rem:symmetrizing},
  we ensure that applying the $b$-dimensional version of
matrix $Q$ to the symmetrized skip-sample DFT (so as to invert the DFT and bring us back to the {\it time domain}) 
will yield a real-valued
  vector of length $b$; this can be useful for statistics that are formulated in the time domain.
However, there is an interesting class of statistics that are defined 
in the frequency domain; three prime examples are discussed in Section
\ref{se:Statistics defined in the frequency domain}.
The next section defines the new skip-sampling methodology for
such statistics defined in the frequency domain.

 \section{Skip-sampling: the basic methodology and some key results} 
\label{se:Skip-sampling}

As before, let $X_1, X_2, \ldots, X_T $ be an observed sample 
from a  strictly stationary time series $\{ X_t \}$ with mean $\mu$
and absolutely summable  autocovariance $ \gamma_k
= \mbox{Cov} (X_0,X_k)$ so that     the spectral density  
$f(\lambda)=\sum_{-\infty}^\infty \gamma_k e^{-ik\lambda}$
is well-defined and continuous on $[-\pi, \pi]$, i.e., belongs to $C[-\pi, \pi]$.
A crude estimate of $f(\lambda)$
is given by the periodogram $I_T(\lambda)=\sum_{-T+1}^{T-1} \hat \gamma_k e^{-ik\lambda}$,
 where the sample autocovariance is defined as 
$\hat \gamma_k =T^{-1} \sum_{t=1}^{T-|k|} (X_t-\bar X)(X_{t+|k|}-\bar X)$,
and $\bar X = T^{-1} \sum_{t=1}^{T} X_t$ is the sample mean.

Interestingly, when evaluated at a (nonzero) Fourier
frequency, the 
periodogram equals the  squared magnitude of the  DFT.  To see that, note the
identity $I_T(\lambda)=T^{-1}|\sum_{t=1}^T (X_t-\bar X)e^{-it\lambda}|^2$.
One of the columns of the matrix $Q$  
consists of constant elements, and the other columns are orthogonal to it. 
Hence, 
\[
I_T(0)=0,  \ \ \mbox{and when} \ \  \ell  \neq 0 \ \ \mbox{we have} \ \
I_T ( \lambda_\ell)=T^{-1}|\sum_{t=1}^T  X_t e^{-it\lambda_\ell}|^2,
\ \ \mbox{where} \ \ \lambda_\ell =   2 \pi \ell/T .
\]

Let $\theta$ be a parameter of interest; we will assume that 
$\theta$ is some real-valued functional   of $f$, i.e., that 
$\theta={\cal G}(f)$ for some    ${\cal G}:C[-\pi, \pi]\to  {\bf R}$. 
The periodogram is asymptotically unbiased but inconsistent for $f(\lambda)$,
as its variance does not tend to zero;  see Chapter 9 of McElroy and Politis (2020). 
However, there are several situations where  
$\theta$ can be consistently estimated using the periodogram as a basis. 
So consider a statistic  $\widehat{\theta}_T$ 
that is a functional of $I_T$, i.e., that
$\widehat{\theta}_T={\cal G}_T (I_T)$ where, for each $T$, we have 
${\cal G}_T:C[-\pi, \pi]\to  {\bf R}$. 
In simpler cases, the functional ${\cal G}_T$ might not 
depend on $T$, as in the case  of spectral means and ratio 
statistics discussed in Sections \ref{se:Spectral means} and \ref{se:Ratio statistics}.

In terms of feasible statistical computing, we will further assume ---as it
is invariably the case--- that the statistic  $\widehat{\theta}_T$ is
computable based on the periodogram evaluated just on the 
Fourier frequencies. Since 
 the periodogram evaluated at (nonzero) Fourier frequencies
equals the  squared magnitude of the DFT, we will assume   that 
\begin{equation}
 \widehat{\theta}_T= {{\cal H}}_T\left( \widetilde{{\bf X}}\right),
\label{eq:general statistic}
\end{equation}
 where   for each $T$, the function $ {{\cal H}}_T$ maps
$  {\bf C}^T$  to $  {\bf R}$.
 We will further assume:
\\

\noindent {\bf Assumption (A)}: For some   nondegenerate limit distribution $J$, we have
$a_T(\widehat{\theta}_T-\theta)\convinlaw J$ as $T\to \infty$, where
  $a_T=T^\delta L(T)$ for some $\delta >0$ and some slowly varying 
function $L$.
\\

\noindent Letting $J_T(x)=Prob\{ a_T(\widehat{\theta}_T-\theta) \leq x\}$, 
Assumption (A) implies that $J_T(x)\to J (x)$ for all   points 
$x$ at which $J$ is continuous.

We can now define the $j$th  skip-sample statistic
\begin{equation}
 \widehat{\theta}_b^{(j)}= {{\cal H}}_b\left( \widetilde{{\bf X}}^{(j)}\right)
\label{eq:general skip-sample statistic}
\end{equation}
for $j=1, \ldots, q$. 
The idea is that $\widehat{\theta}_b^{(j)}$ will have the same 
asymptotic distribution as $\widehat{\theta}_b^{(1)}$ when $b\to \infty$. 
Futhermore, under standard conditions ---see e.g. Lahiri (2003b)---, the DFT ordinates evaluated at different Fourier frequencies will be  asymptotically independent; this would render the
skip-sample statistics $\widehat{\theta}_b^{(1)},\ldots , \widehat{\theta}_b^{(q)}$  (for fixed $q$) 
approximately independent as well. 

We formulate these stylized facts in the following Assumption, which operates under the
condition 
\begin{equation}
\frac {b}{T}  +\frac{1}{b} \to 0 \ \ \mbox{as} \ \ T\to \infty . 
\label{eq:bT}
\end{equation} 
\vskip .1in
\noindent {\bf Assumption (A$^*$)}: Under condition 
(\ref{eq:bT}) the following are true: 
(a) For any $j$, $Prob\{ a_b(\widehat{\theta}_b^{(j)}-\theta)\leq x\}-J_b(x)=o(1)$
 for all   points 
$x$ at which $J$ is continuous; and
(b) for any $j\neq k $, and any bounded functions
$g_1, g_2$,  we have 
$\mbox{Cov} (g_1(\widehat{\theta}_b^{(j)}), g_2(\widehat{\theta}_b^{(k}))\to 0$. 
\\

The quantity $a_T(\widehat{\theta}_T-\theta)$ is sometimes called
a `root'.    
Our core result is a ``subsampling in the frequency domain'' consistency theorem that gives 
conditions under which
the empirical distribution of the skip-sample
roots $ a_b(\widehat{\theta}_b^{(j)}-\widehat{\theta}_T) $
for $j=1, \ldots, q$
can be used to approximate the distribution of the original root.
To develop it,  
define the   two skip-sampling distributions:
$$U_{b,T}(x)=q^{-1} \sum_{j=1}^q {\bf 1}\{a_b(\widehat{\theta}_b^{(j)}-\theta)\leq x \}
\ \ \mbox{and} \ \ L_{b,T}(x)=q^{-1} \sum_{j=1}^q {\bf 1}\{a_b(\widehat{\theta}_b^{(j)}-\widehat{\theta}_T) \leq x \},$$
where  $ {\bf 1}$ denotes the indicator function.
Of the two skip-sampling distributions, 
$U_{b,T}$ is termed an `oracle' as it requires knowledge of $\theta$
for its   construction. By contrast, $
L_{b,T}$ is a {\it bona fide} statistic that can be used for estimation purposes.

\begin{Theorem}
 \label{thm:skip-sample.consistency}
Assume condition (\ref{eq:bT}) and  Assumptions (A) and  (A$^*$).  Then,   
$$ L_{b,T}(x)   \convinp J(x)  $$
 for all   points 
$x$ at which $J$ is continuous; here  $\convinp $ denotes convergence
in probability.
\end{Theorem}
\noindent
{\bf Proof.}  Let $x$ be  a point of continuity of $J$.  By the first argument given in the proof of Theorem 2.2.1 of
Politis et al. (1999),  $U_{b,T}(x) $ and $ L_{b,T}(x) $  are asymptotically
close. So  to prove Theorem \ref{thm:skip-sample.consistency}, it   suffices to 
show that $ U_{b,T}(x)   \convinp J(x)  $.

Note that $E U_{b,T}(x) = q^{-1} \sum_{j=1}^q Prob\{a_b(\widehat{\theta}_b^{(j)}-\theta)\leq x \}= J_b(x) +o(1)$ by Assumption  (A$^*$)
and the Cesaro sums lemma. 
Furthermore, $J_b(x) \to J(x)$ by Assumption  (A) and (\ref{eq:bT}).
Now note that
\[
\mbox{Var} (U_{b,T}(x) )=q^{-2} \sum_{j=1}^q \sum_{k=1}^q
\mbox{Cov}(g_1(\widehat{\theta}_b^{(j)}), g_1(\widehat{\theta}_b^{(k})),
\]
where
$g_1(\widehat{\theta}_b^{(j)})= {\bf 1}\{a_b(\widehat{\theta}_b^{(j)}-\theta)\leq x \}$.  By Assumption  (A$^*$)
and   Cesaro sums, it follows that
\[
\mbox{Var} (U_{b,T}(x) )=q^{-2} \sum_{j=1}^q  
\mbox{Cov} (g_1(\widehat{\theta}_b^{(j)}), g_1(\widehat{\theta}_b^{(j) })) +o(1).
\]
Since $g_1$ is an indicator, it follows that 
$|\mbox{Cov} (g_1(\widehat{\theta}_b^{(j)}), g_1(\widehat{\theta}_b^{(j) }))|\leq 1$. 
Hence, $\mbox{Var} (U_{b,T}(x) )= o(1) $, 
and  the desired  result 
follows by Chebyshev's inequality. $\Box$

\begin{Remark} \label{re:sub variance} \rm
In many situations, the limit law $J$ will be $N(0,v )$. In this case, it may 
be of interest to use skip-sampling to estimate the asymptotic variance
$v $, and use the normal tables (instead of the quantiles of the
skip-sampling distribution $L_{b,T}$) in order to construct confidence intervals and 
tests. The skip-sampling estimator of $v $ is given by
\begin{equation}
 \label{eq:sub variance}
 \widehat v_b=  
\frac{a_b^2}{q} \sum_{j=1}^q ( \widehat{\theta}_b^{(j)} - \bar{\widehat{\theta}}_b)^2,
\end{equation}
where $\bar{\widehat{\theta}}_b= q^{-1} \sum_{j=1}^q \widehat{\theta}_b^{(j)} $. The consistency of $\widehat v_b$ requires some different conditions that
are outlined in Corollary \ref{cor:skip-sample.consistency variance}
 below. Such conditions can be  verified  for the two prominent  
types of periodogram-based statistics, namely spectral means and ratio 
statistics; see   Sections \ref{se:Spectral means} and \ref{se:Ratio statistics}.  
Additional  examples of  potential applicabilty of
frequency domain resampling (including skip-sampling)
are given in Corollary 3.1   of 
Bertail and Dudek (2021) and its related  discussion. 

\end{Remark} 

\begin{Corollary}
 \label{cor:skip-sample.consistency variance}
Assume Assumption (A) with
  $\sup_T E\widehat{\theta}_T^4<\infty$, 
and $a_T^2 \mbox{Var} [ \widehat{\theta}_T ] \to v >0$
as $T\to \infty$.
Let $b$ be a sequence satisfying
\begin{equation}
\frac {b}{T}  +\frac{1}{b} + \frac{a_b^{ 2}}{T} \to 0 \ \ \mbox{as} \ \ T\to \infty . 
\label{eq:bT2}
\end{equation}  
Also assume that, for any $1 \leq i, j \leq q$ and   $i \neq j$,
 the following set of assumptions holds:
 
\begin{equation}
 \begin{cases}
& \EE [ \widehat{\theta}_b^{(j)} ]   =  \theta + o( a_b^{-1})
\\ 
& a_b^2  \mbox{Var} [ \widehat{\theta}_b^{(j)} ]   = 
v+o(1)
\\
& \mbox{Cov} [ \widehat{\theta}_b^{(i)}, \widehat{\theta}_b^{(j)} ]  
  =    O(T^{-1})  .
\end{cases}
 \label{eq:3 assumptions}
\end{equation}
 Further assume that  when $i\neq j$,
\begin{equation}
 \mbox{Cov} \left[a_b^{2} ( \widehat{\theta}_b^{(i)} -  {\theta} )^2   , \ 
a_b^{2} ( \widehat{\theta}_b^{(j)} -  {\theta} )^2   \right]
= o(1).
\label{eq:4th moments}
\end{equation}
 Then, $\widehat v_b \convinp v$ as $T\to \infty$.
\end{Corollary}

\begin{Remark} \rm
\label{rem:corollary 1}  
The validity of assumption (\ref{eq:4th moments})   can be motivated by
   the asymptotic independence of $\widehat{\theta}_b^{(i)}$ and $\widehat{\theta}_b^{(j)}$.
Moreover, note that 
 the set of assumptions (\ref{eq:3 assumptions}) implies
  $ a_b^2E( \widehat{\theta}_b^{(i)} -  {\theta} )^2=v+o(1)$.
Hence, by Markov's inequality,
  $  a_b^2 ( \widehat{\theta}_b^{(i)} -  {\theta} )^2=v+o_P(1)$,
 implying that 
$  \left[a_b^{2} ( \widehat{\theta}_b^{(i)} -  {\theta} )^2 - v \right]\left[ a_b^{2} ( \widehat{\theta}_b^{(j)} -  {\theta} )^2 - v  \right]
$ $= o_P(1)$.
Eq. (\ref{eq:4th moments}) can be be viewed as a stronger version
of this result.
\end{Remark}

\paragraph{Proof of Corollary \ref{cor:skip-sample.consistency variance}. }
Consider the `oracle'  quantity 
$\widetilde v_b=\frac{a_b^2}{q} \sum_{j=1}^q ( \widehat{\theta}_b^{(j)} -  {\theta} )^2,$
and note that 
\begin{equation}
\widehat v_b=  
\frac{a_b^2}{q} \sum_{j=1}^q ( \widehat{\theta}_b^{(j)} -
\theta+\theta - \bar{\widehat{\theta}}_b)^2
= 
\widetilde v_b -a_b^2 (    \bar{\widehat{\theta}}_b-\theta )^2 .
\label{eq:v oracle}
\end{equation} 
By assumption, $ E     \bar{\widehat{\theta}}_b=\theta + o( a_b^{-1})$,
and 
\[
q^2\mbox{Var}[\bar{\widehat{\theta}}_b]=
\sum_{i=1}^q  \sum_{j=1}^q 
 \mbox{Cov} [ \widehat{\theta}_b^{(i)}, \widehat{\theta}_b^{(j)} ]
=
\sum_{i=1}^q \mbox{Var}[\widehat{\theta}_b^{(i)}]
+ \sum_{i=1}^q  \sum_{j\neq i}  
 \mbox{Cov} [ \widehat{\theta}_b^{(i)}, \widehat{\theta}_b^{(j)} ]
= q   a_b^{-2} (v+o(1)) + O(\frac{q^2}{T})
\]
so that 
\[
a_b^{ 2}E(    \bar{\widehat{\theta}}_b-\theta )^2
= \frac{ v+o(1)}{q     }+ O(\frac{a_b^{ 2}}{T}), 
\]
which tends to zero by (\ref{eq:bT2}). 
Hence, eq.~(\ref{eq:v oracle}) implies that  
$\widehat v_b-\widetilde v_b  \convinp 0$. 

To show $\widetilde v_b \convinp v$, first note that assumption 
 (\ref{eq:3 assumptions}) implies 
$$
E\widetilde v_b = \frac{a_b^{ 2}}{q}
\sum_{j=1}^q 
\left(\mbox{Var} [\widehat{\theta}_b^{(j)}]
+\mbox{Bias}^2 [\widehat{\theta}_b^{(j)}]\right)
=v+o(1), $$ 
  i.e., $\mbox{Bias}[\widetilde v_b]\to 0$. 
 To show $\mbox{Var} [\widetilde v_b]\to 0$, 
note that
$$\widetilde v_b-v=\frac{a_b^2}{q} \sum_{j=1}^q ( \widehat{\theta}_b^{(j)} -  {\theta} )^2 -v =
  \frac{1}{q} \sum_{j=1}^q s_j,
$$
letting  $s_j= a_b^{2}( \widehat{\theta}_b^{(j)} -  {\theta} )^2 - v$. 
Finally,   
$$ E(\widetilde v_b-v)^2   
=\frac{1}{q^2}  \sum_{i=1}^q \sum_{j=1}^q
E (s_i s_j)
=\frac{1}{q^2}  \sum_{i=1}^q E s_i^2
+\frac{1}{q^2}  \sum_{i=1}^q \sum_{j\neq i}
E ( s_i s_j) = O(1/q)+o(1)
$$
by (\ref{eq:4th moments}).  
By Chebyshev's inequality, it follows that
 $\widetilde v_b \convinp v$, 
and therefore $\widehat v_b \convinp v$ as well.
$\Box$

\section{Examples of skip-sampling applicability} 
\label{se:Statistics defined in the frequency domain}
 \subsection{Spectral means} \label{se:Spectral means}
Consider a bounded function  $g   (\lambda) $ of   domain $[-\pi,\pi ]$
 that has  bounded variation, and denote 
$\langle g  \rangle = { ( 2 \pi )}^{-1} \, \int_{-\pi}^{\pi} g(\lambda)  \, d\lambda. $
A   linear {\it spectral mean}  (Dahlhaus, 1985) is a parameter of the form
\begin{equation}
  \theta = \langle g \, f \rangle = \frac{1}{ 2 \pi} \, \int_{-\pi}^{\pi} g(\lambda) \, f(\lambda) \, d\lambda.
\label{eq.spectralmean}
\end{equation}
The prime example of a linear spectral mean is the 
autocovariance at lag $k$, where $g(\lambda)=e^{ik\lambda}$.

As already mentioned, the periodogram is asymptotically unbiased but inconsistent for $f(\lambda)$,
as its variance does not tend to zero.
However, plugging in $I_T$ instead of $f$ in eq. (\ref{eq.spectralmean})
yields a consistent estimator of the  spectral mean, since
 integration works like summation in terms of reducing the variance.  
As a matter of fact, the integral in eq. (\ref{eq.spectralmean})
is typically approximated by a Riemann sum over the Fourier
frequencies.  Consequently,  a linear spectral mean $\theta$ satisfying eq.~(\ref{eq.spectralmean}) 
is practically estimated by  
\[
  \widehat{\theta}_T =  T^{-1} \, 
\sum_{\ell \in R_T } 
g( 2 \pi \ell/T) \,   I_T ( 2 \pi \ell/T),
\]
 where we define the index range $ R_T =\{\ell :  [ T/2] - T +1 \leq \ell \leq [T/2] \}$. 
%

To prove our next result, we
  consider stationary non-Gaussian processes that satisfy autocumulant conditions described in Taniguchi and Kakizawa (2000).
  Supposing that all moments exist and the autocumulant functions are deﬁned via
\[
  \gamma_{h_1, \ldots, h_{k-1}} = \mbox{cum} \{ X_{t+h_1}, X_{t+h_2}, \ldots, X_{t+h_{k-1}}, X_t \},
\]
we will entertain:
\\

\noindent {\bf Assumption (B)}: for all $k \geq 2$  and each $j = 1, \ldots, k-1$   we have 
 \[
 \sum_{ h_1  \in \ZZ } 
\cdots
\sum_{  h_{k-1} \in \ZZ }  
(1 + |h_j|) \, | \gamma_{h_1, \ldots, h_{k-1}} | < \infty.
\]
\vskip .1in
\noindent
For a process satisfying Assumption (B) the tri-spectral density $F$ is well-defined, and is given by
 \[
  F(\omega_1, \omega_2, \omega_3) = \sum_{h_1 \in \ZZ}
\sum_{h_2 \in \ZZ} \sum_{h_3 \in \ZZ}  \gamma_{h_1, h_2, h_3}
  \exp \{ -i (h_1 \omega_1 + h_2 \omega_2 + h_3 \omega_3 ) \},
  \]
in which case 
  it can be shown (McElroy and Roy, 2022) that
\begin{equation}
\label{eq:CLT}
\sqrt{T} \, \left( \widehat{\theta}_T - \theta \right) \convinlaw \mathcal{N} \left( 0,  \langle g \,    g^{\star}  \, f^2 \rangle
	+   \langle \langle g \, g \,  F \rangle \rangle \right)
\ \ \mbox{as} \ \ T \to \infty ;
\end{equation}
here, we have used the short-hand
$g^{\star}  (\lambda) = g(\lambda) + g^{\sharp} (\lambda),$
where $g^{\sharp}$ is the reflection of $g$ about the y-axis, and we have denoted
\[
  \langle \langle g \, g \,  F \rangle \rangle = \frac{1}{ {(2 \pi)}^2 } \, \int_{-\pi}^{\pi} \int_{-\pi}^{\pi}
    g(\lambda) \, g(\omega) \, F(\lambda, - \lambda, \omega) \, d\lambda d\omega.
\]
 Dahlhaus    (1985)  proved eq. (\ref{eq:CLT}) for linear processes.
Working under Assumption (B) allows us to go beyond the 
setting of linearity; however, there could be different sufficient conditions
for (\ref{eq:CLT}).

 Evaluating  the linear spectral mean on the $j$th skip-sample DFT yields
\begin{equation}
\label{eq:first-def}
   b^{-1} \, \sum_{\ell \in R_b } g( 2 \pi \ell/b) \, I_T ( 2 \pi \ell/b + 2 \pi j/T).
\end{equation}
The displacement of the periodogram by $2 \pi j/T$ means that
 this is no longer an even function of $ 2 \pi \ell/b$ (unless $j = q$), so the asymptotic variance 
 has no contribution from $g^{\sharp}$.  In order to correct this, we need to define skip-samples
 over $[0, \pi]$ and then reflect onto $[-\pi,0]$, which suggests the   definition of
 the $j$th {\em skip-sample statistic}
\[
 \widehat{\theta}_b^{(j)} 
	= b^{-1} \, \sum_{\ell =1}^{[b/2]} g^{\star} ( 2 \pi \ell/b) \, I_T ( 2 \pi \ell/b + 2 \pi j/T).
\]  
The above is obtained from (\ref{eq:first-def})
 by imposing that at Fourier frequencies $ 2 \pi \ell/b$ with negative $\ell$, we evaluate the
 periodogram at $ 2 \pi \ell/b - 2 \pi j/T$ instead of at $ 2 \pi \ell/b + 2 \pi j /T$.  Then,
 ignoring the contribution from $\ell =0$, we obtain the above expression
 for $\widehat{\theta}_b^{(j)}$  in terms of $g^{\star} (\lambda)$.
 Next, we summarize some of the moment properties of these skip-sample statistics.

\begin{Theorem}
 \label{thm:skip-sample.clt}
 Assume that $\{ X_t \}$ is strictly stationary and satisfies Assumption B. 
Consider a linear spectral mean $\theta$ satisfying eq. (\ref{eq.spectralmean}), 
and some fixed bounded function      $g   (\lambda) $
having  bounded variation.  Let $b$ be a sequence satisfying
(\ref{eq:bT}). Then, for any $1 \leq j \leq q$, we have 
\begin{align*}
 \EE [ \widehat{\theta}_b^{(j)} ] & =  \theta +  O(b^{-1}) + O(T^{-1}) \\
  \mbox{Var} [ \widehat{\theta}_b^{(j)} ] & =  b^{-1} \, \langle g \, g^{\star}  \, f^2 \rangle
  + T^{-1} \,  \langle \langle g \, g \,  F \rangle \rangle  + O(T^{-2})  + O(b^{-2}) +  O (b^{-1} T^{-1}).
\end{align*}
 Also, for $1 \leq i,j \leq q$ and  $i \neq j$, we have 
$
 \mbox{Cov} [ \widehat{\theta}_b^{(i)}, \widehat{\theta}_b^{(j)} ]  =    O(T^{-1}).
$
\end{Theorem}

\paragraph{Proof of Theorem \ref{thm:skip-sample.clt}.}  
Let $d_T (\lambda) = \sum_{t=1}^T X_t e^{-i t \lambda }$,  so that for $\lambda \neq 0$ we have
$I_T (\lambda) = T^{-1} {| d_T (\lambda) |}^2$.   When evaluated at a non-zero Fourier frequency $\lambda_{\ell}$,
$d_T (\lambda_{\ell})$ gives the same value when computed from $X_t - \mu$ instead of $X_t$,
because $\sum_{t=1}^T e^{-i t \lambda_{\ell}} = 0$.  So without loss of generality we can assume that
$\mu = 0$ in our analysis.  Furthermore, because 
$\EE [ d_T (\lambda) d_T (-\lambda) ] = \mbox{cum} ( d_T (\lambda), d_T (-\lambda)) $ when $\mu = 0$,
 we can apply Theorem 4.3.2 of Brillinger (1981)  to obtain
 \begin{align*}
 \EE [ \widehat{\theta}_b^{(j)} ] & = \frac{1}{bT} \sum_{\ell =1}^{[b/2]} 
  \left( g( 2 \pi \ell/b) + g (-2 \pi \ell/b) \right)   \,  \EE [ d_T ( 2 \pi \ell/b + 2 \pi j/T)
   d_T ( -2 \pi \ell/b - 2 \pi j/T) ] \\
   & =  \frac{1}{bT} \sum_{\ell =1}^{[b/2]} 
  \left( g( 2 \pi \ell/b) + g (-2 \pi \ell/b) \right)   \,   \left( 
  T f( 2 \pi \ell/b  + 2 \pi j /T) + O(1)  \right) \\
  & = O(T^{-1}) +  \frac{1}{b} \sum_{\ell =1}^{[b/2]} 
  \left( g( 2 \pi \ell/b) + g (-2 \pi \ell/b) \right)   \,   
   f( 2 \pi \ell/b  + 2 \pi j /T),
 \end{align*}
using the boundedness of $g$.  The case $k=2$ of Assumption (B) implies that $\partial_\lambda f(\lambda)$ is bounded
in $\lambda$, and hence by a Taylor series expansion $ f( 2 \pi \ell/b  + 2 \pi j /T)
= f ( 2 \pi \ell/b) + O(b^{-1})$, since $|j|/T \leq q/T = b^{-1}$.  Thus,
$ \EE [ \widehat{\theta}_b^{(j)} ]$ equals  $ b^{-1} \sum_{\ell =1}^{[b/2]} 
  \left( g( 2 \pi \ell/b) + g (-2 \pi \ell/b) \right)   \,      f( 2 \pi \ell/b )$ plus terms
  that are $O(T^{-1}) + O(b^{-1})$.  Finally,
  \[
  b^{-1} \sum_{\ell =1}^{[b/2]} 
  \left( g( 2 \pi \ell/b) + g (-2 \pi \ell/b) \right)   \,      f( 2 \pi \ell/b )
  = b^{-1} \sum_{|\ell| = 1}^{[b/2]} g(2 \pi \ell/b) \, f( 2 \pi \ell/b)
\]
by the evenness of $f$, and this last expression is the Riemann sum on a mesh of size $b^{-1}$ of
 $\langle g f \rangle = \theta$.  This proves the first assertion.  
 
 For the variance, we write
 \begin{align*}
  \mbox{Var} [ \widehat{\theta}_b^{(j)} ] 
 &  = \frac{1}{b^2 T^2}  \sum_{\ell,k=1}^{[b/2]} 
  \left(  g ( 2 \pi \ell/b ) + g ( - 2\pi \ell/b) \right) 
  \left( g(2 \pi k/b) + g (- 2 \pi k/b) \right) \\
  & \cdot  \mbox{cum} \left( d_T (2 \pi \ell/b + 2 \pi j/T) d_T (-2 \pi \ell/b - 2 \pi j/T),
      d_T (2 \pi k/b + 2 \pi j/T) d_T (-2 \pi k/b - 2 \pi j/T) \right).
\end{align*}
To compute the cumulant, we use Theorems 2.3.2 and 4.3.2 of Brillinger (1981).
In the case that $\ell = k$, we find the cumulant, up to terms that are $O(1)$, is
\[
  T^2 { f( 2 \pi \ell/b + 2 \pi j/T) }^2 +  T F(  2 \pi \ell/b + 2 \pi j/T, 
    - 2 \pi \ell/b - 2 \pi j/T,  2 \pi \ell/b + 2 \pi j/T ).
\]
Hence we obtain the ``diagonal'' contribution to $ \mbox{Var} [ \widehat{\theta}_b^{(j)} ] $ is
\begin{align*}
&  \frac{1}{b^2 T^2}  \sum_{\ell=1}^{[b/2]} 
  {\left(  g ( 2 \pi \ell/b ) + g ( - 2\pi \ell/b) \right) }^2 \\
 & \cdot  \left(  T^2 { f( 2 \pi \ell/b + 2 \pi j/T) }^2 +  T F(  2 \pi \ell/b + 2 \pi j/T, 
    - 2 \pi \ell/b - 2 \pi j/T,  2 \pi \ell/b + 2 \pi j/T ) \right),
\end{align*}
 and the term involving the tri-spectrum is of lower order, and therefore is vanishing 
 asymptotically.  So the   the diagonal contribution is asymptotic to
 \[
  \frac{1}{2 \pi b}  \int_0^{\pi} { (g (\lambda) + g(-\lambda) ) }^2 \, { f(\lambda) }^2 d\lambda
  =    \frac{1}{2 \pi b}  \int_{-\pi}^{\pi} g (\lambda) (g (\lambda) + g(-\lambda) )  \, { f(\lambda) }^2 d\lambda
 = b^{-1} \langle g (g + g^{\sharp}) f^2 \rangle.  
  \]
  Turning to the ``off-diagonal'' contribution to the variance, we examine the case that $\ell \neq k$, finding
  that the cumulant is
  \[
  T F(  2 \pi \ell/b + 2 \pi j/T,     - 2 \pi \ell/b - 2 \pi j/T,  2 \pi k/b + 2 \pi j/T )
  \]
up to terms that are $O(1)$.  We can utilize a Taylor series expansion in each of the three arguments of the 
tri-spectrum, finding that up to terms $O(b^{-1})$ we have 
$F(  2 \pi \ell/b,     - 2 \pi \ell/b,  2 \pi k/b )$.  Hence the off-diagonal portion of the asymptotic variance is
\begin{align*}
&   \frac{1}{b^2 T}  \sum_{\ell \neq k=1}^{[b/2]} 
  \left(  g ( 2 \pi \ell/b ) + g ( - 2\pi \ell/b) \right) 
  \left( g(2 \pi k/b) + g (- 2 \pi k/b) \right) \, F(  2 \pi \ell/b,     - 2 \pi \ell/b,  2 \pi k/b ) \\
   &  =  \frac{ {(b/2)}^2 }{ b^2 T \pi^2}  \int_0^{\pi} \int_0^{\pi}
    ( g (\lambda) + g(-\lambda) ) ( g (\omega) + g(- \omega) ) \,
      F(\lambda, -\lambda, \omega) \, d\lambda d\omega  + O (b^{-1} T^{-1}),
\end{align*}      
      using the Riemann sum approximation for both integrals.  Exploiting the symmetries of
      the tri-spectrum, the above expression simplifies to $ T^{-1} \langle \langle  g g F \rangle \rangle
      + O (b^{-1} T^{-1})$.
      
      Finally,   the covariance (for $i \neq j$) is
\begin{align*}
 \mbox{Cov} [ \widehat{\theta}_b^{(i)},  \widehat{\theta}_b^{(j)}  ] 
  &  = \frac{1}{b^2 T^2}  \sum_{\ell,k=1}^{[b/2]} 
  \left(  g ( 2 \pi \ell/b ) + g ( - 2\pi \ell/b) \right) 
  \left( g(2 \pi k/b) + g (- 2 \pi k/b) \right) \\
  & \cdot  \mbox{cum} \left( d_T (2 \pi \ell/b + 2 \pi i/T) d_T (-2 \pi \ell/b - 2 \pi i/T),
      d_T (2 \pi k/b + 2 \pi j/T) d_T (-2 \pi k/b - 2 \pi j/T) \right).
\end{align*}
 First examining the case that $\ell = k$,  the cumulant 
 involves terms that are $O(1)$,  plus
 \[
     T^2  f( 2 \pi \ell/b + 2 \pi i/T)   f( 2 \pi \ell/b + 2 \pi j/T)  
{\bf 1}_{\{i = j\}}
    + T F(  2 \pi \ell/b + 2 \pi i/T, 
    - 2 \pi \ell/b - 2 \pi i/T,  2 \pi \ell/b + 2 \pi j/T ).
    \]
The indicator arises, because in the cumulant asymptotics of Theorem 4.3.2 of Brillinger (1981), 
 there is a Dirac operator evaluated at the sum of Fourier frequencies 
  $ -2 \pi \ell/b - 2 \pi i/T + 2 \pi \ell/b + 2 \pi j/T$, which is zero if and only if $i =j$.  Since
  this case is excluded by assumption, we find that the diagonal contribution to the covariance is
  $O(T^{-1})$. 
For the case that $\ell \neq k$,  we now consider a Dirac operator evaluated at the sum of
Fourier frequencies  $ -2 \pi \ell/b - 2 \pi i/T + 2 \pi k/b + 2 \pi j/T$, which equals zero if and only
 if $j-i = q (\ell -k)$.  But since $i \neq j$, we see that $|j-i| \in \{1, \ldots, q-1 \}$, and so the
  condition is impossible.   Hence the only contribution to the cumulant is the tri-spectrum,
   and we obtain 
 \[
     O(1)    + T F(  2 \pi \ell/b + 2 \pi i/T, 
    - 2 \pi \ell/b - 2 \pi i/T,  2 \pi k/b + 2 \pi j/T ).
    \]
It follows that the covariance is $O(T^{-1}$) asymptotically, which completes the proof.  $\quad \Box$

\vspace{.5cm}

 Denote  
the left-hand-side of (\ref{eq:CLT}) by  $S_T (\theta)=\sqrt{T} \, \left( \widehat{\theta}_T - \theta \right)$; this was previously called a `root'.
As a consequence of Theorem
 \ref{thm:skip-sample.clt}, the $j$th  oracle skip-sample root
$  S^{(j)}_T (\theta) = \sqrt{ b} \, ( \widehat{\theta}_b^{(j)} - \theta)$
 has asymptotic variance  
\begin{equation}
\label{eq: problem with variance}
\langle g \, g^{\star}  \, f^2 \rangle  + (b/T) \,  \langle \langle g \, g \,  F \rangle \rangle.
\end{equation}
Theorem  \ref{thm:skip-sample.clt} allows us to take advantage of 
the avenue suggested by Remark \ref{re:sub variance}, i.e., using the 
skip-sampling estimator of the asymptotic variance 
of eq. (\ref{eq:CLT}), and then use the normal tables for inference.
Note that here the rate $a_T=\sqrt{T}.$
Letting $b=o(\sqrt{T})$  to satisfy (\ref{eq:bT2}), 
Corollary \ref{cor:skip-sample.consistency variance}
would be applicable as long as assumption
(\ref{eq:4th moments}) were also verified.  
In view of Remark \ref{rem:corollary 1}, the latter is   
expected to hold but it is cumbersome to evaluate. 

However, there is an additional issue: 
since $b/T \tends 0$, the second term of
(\ref{eq: problem with variance}) asymptotically drops out,
which  is undesirable in terms of capturing
the variance given in (\ref{eq:CLT}).
To elaborate, for roots such that
 $\langle \langle g \, g \,  F \rangle \rangle = 0$ the asymptotic variance of 
 $S_T^{(j)} (\theta)$ is correct, but otherwise must be adjusted to account 
 for the non-trivial contribution from the tri-spectrum.
  Whenever $\langle \langle g \, g \,  F \rangle \rangle = 0$, we
will say the asymptotic 
 distribution (\ref{eq:CLT}) is ``tri-spectrum free."

 \begin{Corollary}
 \label{co:spectral means}
 Assume the assumptions of Theorem  \ref{thm:skip-sample.clt},  
the additional assumption  (\ref{eq:4th moments}) and $b=o(\sqrt{T})$.
Then,  the skip-sampling estimator $\hat v_b$   from eq. 
 (\ref{eq:sub variance}) 
 is consistent for the asymptotic variance 
 appearing in eq. (\ref{eq:CLT}) when the latter is ``tri-spectrum free."
 \end{Corollary}

\begin{Remark} \rm
\label{rem:tri-spectrum.free}
Requiring that the asymptotic distribution be  ``tri-spectrum free" is  common with several 
resampling methods in the frequency domain.
For example,  the original   frequency domain bootstrap of Franke and H\"ardle (1992) fails to capture the second term 
of (\ref{eq: problem with variance}) even for linear processes;
 see Paparoditis (2002) for a review.
 In general, by the Wold decomposition  we have $X_t =EX_0+  \sum_{j \geq 0} \psi_j \epsilon_{t-j}$, where the sequence  $\epsilon_t $ is mean zero, uncorrelated
with variance $\sigma^2$, i.e., a white noise,
   but  not
necessarily independent, identically distributed (i.i.d.).  If the process
$\{X_t\}$ is linear (and causal), then 
 $\epsilon_t \sim \mbox{i.i.d.}$ as well, and the expression for the variance greatly simplifies.  In this case, 
\[
\langle \langle g \, g \,  F \rangle \rangle = (\eta - 3) \, { \langle g \, f \rangle }^2,
\]
 where $\eta= \EE [ \epsilon_t^4]/\sigma^4$.   This yields a classical result: for linear time series, if the innovation kurtosis is that of a Gaussian
  (i.e., $\eta = 3$),
  or in the special case when
the linear spectral mean is zero (i.e., $\langle g \, f \rangle =0$),
 then the asymptotic distribution (\ref{eq:CLT}) is tri-spectrum free
and Corollary \ref{co:spectral means} is applicable. 
\end{Remark}

If  $\{X_t\}$ is linear but $\eta \neq 3$, it may still be possible to conduct inference
on spectral means via a hybrid procedure employing skip-sampling as
a component. For example, let 
$\widehat \eta$ be the estimator of  $\eta$ based on 
  the technique of Fragkeskou and Paparoditis (2016), and let 
$\widehat f$ be a consistent estimator of the spectral density $f$.
Then, we can   estimate the asymptotic variance 
appearing in eq. (\ref{eq:CLT}) by 
$\widehat v_b + (\widehat \eta - 3) \, { \langle g \, \widehat f \rangle }^2.$
Alternative hybrid methods are also available, see e.g.    Janas and Dahlhaus (1994), 
Kreiss and Paparoditis (2003), or Meyer et al. (2020).

\subsection{Ratio statistics} 
\label{se:Ratio statistics} 
Consider a parameter $\theta$  that is obtained as the
 finite ratio of two  linear spectral means, i.e.,
$ \theta = \ \langle p \,  f  \rangle / \langle m \, f   \rangle $
 for some fixed bounded functions      $p   (\lambda) $ and $m(\lambda)$ 
having  bounded variation  on $[-\pi, \pi]$. 
We can estimate $\theta$ by a  so-called {\it ratio statistic} 
given by
\begin{equation}
\label{eq:ratio stat}
  \widehat{\theta}_T =  
\frac{ \sum_{\ell \in R_T } p( 2 \pi \ell/T) \,   I_T ( 2 \pi \ell/T)} 
{ \sum_{\ell \in R_T } m( 2 \pi \ell/T) \, I_T( 2 \pi \ell/T)  }.
\end{equation} 
The prime example of a ratio statistic is the sample autocorrelation 
at lag $k$, where $p(\lambda)=e^{ik\lambda}$ and $m(\lambda ) = 1$.

Ratio statistics have an asymptotic distribution that can be tri-spectrum free
under some conditions ---such as linearity of the time series---
and are thus amenable to frequency domain resampling. In fact, 
Dahlhaus  and Janas  (1996) showed that the original frequency domain bootstrap 
of Franke  and H\"ardle   (1992) is not only consistent, but higher-order
accurate for ratio statistics from linear time series. 
To elaborate, Remark \ref{rem:tri-spectrum.free}
implies that  ---provided the process is linear--- 
ratio statistics satisfy a simplified version of 
  (\ref{eq:CLT}), namely 
\begin{equation}
\sqrt{T} \, \left( \widehat{\theta}_T - \theta \right) \convinlaw \mathcal{N} \left( 0,  \langle g \,  g^{\star} \, f^2 \rangle
	/ { \langle m \, f \rangle }^2 \right)
\ \ \mbox{as} \ \ T \to \infty ;
\label{eq:CLT2}
\end{equation}
   here, $g = p - m \theta$, and the 
notation $g^{\star}$ was defined right after (\ref{eq:CLT}).
 In analogy with the previous subsection, we define the $j$th  skip-sample ratio statistic  via
  \[
  \widehat{\theta}_b^{(j)} = \frac{ b^{-1} \sum_{\ell=1}^{[b/2]} p^{\star} (2 \pi \ell/b) \,
   I_T (2 \pi \ell/b + 2 \pi j/T) }{  b^{-1} \sum_{\ell=1}^{[b/2]}  m^{\star} (2 \pi \ell/b) \,
    I_T (2 \pi \ell/b + 2 \pi j/T) }.
 \]  
\begin{Theorem}
 \label{thm:skip-sample-ratio}
 Assume that $\{ X_t \}$ is a strictly stationary linear process that  satisfies Assumption B. 
Consider a finite ratio of  linear spectral means $\theta =  \langle p \,  f  \rangle / \langle m \, f   \rangle$,
 for some fixed bounded functions      $p   (\lambda) $ and $m(\lambda)$ 
having  bounded variation, and let 
  $ \widehat{\theta}_T$ be the ratio statistic
(\ref{eq:ratio stat}).
  Let $b$ be a sequence satisfying
(\ref{eq:bT}), and set  $g = p - m \theta$.   Then, for any $1 \leq j \leq q$, we have 
\begin{align*}
 \EE [ \widehat{\theta}_b^{(j)} ] & =  \theta +  O(b^{-1}) + O(T^{-1}) \\
  \mbox{Var} [ \widehat{\theta}_b^{(j)} ] & =  b^{-1} \, \langle g \, g^{\star}  \, f^2 \rangle/ {\langle m \, f \rangle }^2
    + O(T^{-2})  + O(b^{-2}) +  O (b^{-1} T^{-1}).
\end{align*}
 Also, for $i \neq j$ and $1 \leq i,j \leq q$, we have 
$
 \mbox{Cov} [ \widehat{\theta}_b^{(i)}, \widehat{\theta}_b^{(j)} ]  =    O(T^{-1}).
$
\end{Theorem}

\paragraph{Proof of Theorem  \ref{thm:skip-sample-ratio}.}
It is easy to show that
\[
  \widehat{\theta}_b^{(j)} - \theta = 
  \frac{ b^{-1} \sum_{\ell=1}^{[b/2]} g^{\star} (2 \pi \ell/b) \,
   I_T (2 \pi \ell/b + 2 \pi j/T) }{  b^{-1} \sum_{\ell=1}^{[b/2]}  m^{\star} (2 \pi \ell/b) \,
    I_T (2 \pi \ell/b + 2 \pi j/T) }.
 \]
The denominator, denoted by $\widehat{ \langle m \, f \rangle }$ for short,
 converges in probability to $\langle m \, f \rangle $ by Theorem \ref{thm:skip-sample.clt}.
Because $\theta$ is finite, this limit must be non-zero.  
 Letting the numerator be denoted $\widehat{ \langle g \, f \rangle}$, we obtain
\[
 \widehat{\theta}_b^{(j)} - \theta = 
  \frac{  \widehat{ \langle g \, f \rangle } }{ \langle m \, f \rangle }
   + \frac{    \widehat{ \langle g \, f \rangle }   \left( \langle m \, f \rangle - 
    \widehat{ \langle m \, f \rangle }       \right) }{  \langle m \, f \rangle \,
     \widehat{ \langle m \, f \rangle }   },
\]
and the second term has mean that is $O (b^{-1}) + O (T^{-1}) $, using the Cauchy-Schwarz
inequality, the delta method, and the variance results of    Theorem \ref{thm:skip-sample.clt}.
    This is because the mean of $  \widehat{ \langle g \, f \rangle }$ is zero plus
   lower order terms.  Finally, we can compute the mean
    and variance of $ \widehat{ \langle g \, f \rangle }/ \langle m \, f \rangle$ using
      Theorem \ref{thm:skip-sample.clt}, and obtain the stated results, noting
      that $\langle \langle g g F \rangle \rangle = 0$ since the process is linear and $\langle g \,f \rangle = 0$. $\Box$
  \\
 
Theorem  \ref{thm:skip-sample-ratio} confirms  the validity of 
assumption set (\ref{eq:3 assumptions}) in the context of
ratio statistics; the following corollary then ensues.

\begin{Corollary}
\label{co:ratio statistics}
Assume the assumptions of  Theorem  \ref{thm:skip-sample-ratio}, 
  the additional assumption  (\ref{eq:4th moments}) and
$b=o(\sqrt{T})$. Then, 
the skip-sampling estimator $\hat v_b$   from 
eq.~(\ref{eq:sub variance})
is consistent for the asymptotic variance 
appearing in eq. (\ref{eq:CLT2}).
\end{Corollary}


 


\section*{Acknowledgments}

This report is released to inform interested parties of research and to encourage discussion.  The views expressed on
statistical issues are those of the authors and not  those of the U.S. Census Bureau.  Research of the second
 author partially   supported by NSF grant DMS 19-14556.


\begin{thebibliography}{99}

\bibitem{brill} Brillinger, D.R. (1981).
  {\it Time Series: Data Analysis and Theory}, Holden-Day, New York.
 
   \bibitem{} Bertail, P. and Dudek, A.E. (2021). 
Consistency of the frequency domain
bootstrap for differentiable functionals, 
{\it Electronic Journal of Statistics} {\bf 15}, 1--36.
 

   \bibitem{} Brockwell, P. J.  and Davis, R. A.
  (1991). {\it Time Series: Theory and Methods, 2nd ed.},
Springer, New York.                

\bibitem{} Dahlhaus, R. (1985). 
 Asymptotic normality of spectral estimates, {\it Journal of Multivariate
Analysis} {\bf 16}, 412--431.

\bibitem{} Dahlhaus, R. and Janas, D. (1996). 
A frequency domain bootstrap for ratio statistics in time series analysis,
{\it Ann. Statist.} {\bf 24}(5), 1934--1963.

\bibitem{}    Efron, B. (1979). Bootstrap methods: another look at the jackknife, {\it Ann. Statist.} {\bf 7}, 1--26.

\bibitem{} Franke, J. and H\"ardle, W.  (1992). On bootstrapping 
kernel spectral estimates, {\it Ann. Statist.} {\bf 20}, 121--145.

\bibitem{}  Fragkeskou, M. and Paparoditis, E. (2016).
  Inference for the fourth‐order innovation cumulant in linear time series,
 {\it Journal of Time Series Analysis} {\bf 2}(37), 240--266.

 \bibitem{}  Hurvich, C. M. and Zeger, S. L. (1987). Frequency domain bootstrap methods for time series,
  New York University Working Paper.

 \bibitem{} Janas, D. and R. Dahlhaus (1994). A frequency domain bootstrap for time series. In: {\it Computationally Intensive Statistical Methods. Proceedings of the 26th Symposium on the Interface},  (J. Sall and A. Lehman, eds.). Interface Foundation of North America, Faifax Station, VA, pp. 423--425

\bibitem{}  Kirch, C. and D.N. Politis (2011). TFT-bootstrap: resampling time series 
in the frequency domain to obtain replicates in the time domain, {\it Ann. Statist.} {\bf  39}(3),  1427--1470.

\bibitem{}  Kreiss, J.-P. and   Paparoditis, E. (2003).  Autoregressive aided periodogram bootstrap
for time series, {\it Ann. Statist.}  {\bf 31}(6), 1923--1955.

\bibitem{}  Kreiss, J.-P. and Paparoditis, E. (2023). 
{\it Bootstrap for Time Series: Theory and Applications}, Springer, Heidelberg.

\bibitem{}  Lahiri, S.N. (2003a). {\it Resampling Methods for Dependent Data}, Springer, New York. 

\bibitem{}  Lahiri, S.N (2003b). A necessary and sufficient condition for asymptotic independence 
of discrete Fourier transforms under short-and long-range dependence, {\it Ann. Statist.} {\bf 31}(2), 613--641.

\bibitem{}   McElroy, T.S. and   Politis, D.N. (2020). 
{\it Time Series: A First Course with Bootstrap Starter}, Chapman and Hall/CRC Press,  Boca Raton.

\bibitem{}   McElroy, T.S. and   Roy, A. (2022).   Model identification via total Frobenius norm
 of multivariate spectra,  {\it Journal of the Royal Statistical Society, Series B} {\bf 84}, 473--495.

\bibitem{}  Meyer, M.    Paparoditis, E. and  Kreiss, J.-P. (2020).  
Extending the validity of frequency domain bootstrap methods to general stationary
processes, {\it Ann. Statist.} {\bf 48}(4), 2404--2427.

\bibitem{}  
Paparoditis, E. (2002). Frequency domain bootstrap for time series. In {\it 
Empirical Process Techniques for Dependent Data} (H. Dehling et al., eds.)
365--381. Birkh\"auser, Boston.

\bibitem{}  
Paparoditis, E.  and   Politis, D.N. (1999).  The local bootstrap for periodogram statistics,
{\it Journal of Time Series Analysis} {\bf 20}, 193--222. 

\bibitem{} Politis, D.N.,   Romano, J.P. and Wolf, M. (1999).
  {\it Subsampling}, Springer, New York.


\bibitem{} Taniguchi, M. and Kakizawa, Y. (2000).  {\it Asymptotic Theory of Statistical Inference for
 Time Series},   Springer, New York.

\end{thebibliography}
\end{document}